\documentclass[12pt,preprint]{aastex}
\def\h1{$^1$H}
\def\o16{$^{16}$O}
\def\n14{$^{14}$N}

\def\he4{$^4$He}

\def\bd{\mbox{BD+17 3248}}
\def\cstwo{\mbox{CS 22892-052}}
\def\hdone{\mbox{HD 115444}}
\def\hdtwo{\mbox{HD 221170}}
\def\etal{\mbox{\rm et al.}}

\everymath{\rm}
\everydisplay{\sf}
\received{}
\revised{}
\accepted{}
\slugcomment{To appear in {\it The Astrophysical Journal}}
\shortauthors{Henry et al.}
\shorttitle{Stellar Yields of Fe-Peak Elements}

\begin{document}

\title{EMPIRICALLY DERIVED INTEGRATED STELLAR YIELDS OF Fe-PEAK ELEMENTS} 

\author{R.B.C. Henry$^1$, John J. Cowan$^1$, Jennifer Sobeck$^2$}

\affil{$^1$H.L. Dodge Department of Physics \& Astronomy,
University of Oklahoma, Norman, OK 73019, henry@nhn.ou.edu, cowan@nhn.ou.edu;
$^2$Department of Astronomy and
Astrophysics, University of Chicago, Chicago, IL 60637, jsobeck@uchicago.edu}

\begin{abstract}

We present here the initial results of a new study of massive star yields of Fe-peak elements.
We have compiled from the literature a database of carefully determined
solar neighborhood stellar abundances of seven iron-peak elements, Ti, V, Cr, Mn, Fe, Co, and
Ni, and then plotted [X/Fe] versus [Fe/H] to study the trends as functions
of metallicity. Chemical evolution models were then employed to force a fit to
the observed trends by adjusting the input massive star metallicity-sensitive yields of \citet{Kobay06}.
Our results suggest that yields of Ti, V, and Co are generally larger as well as  anticorrelated with metallicity, in contrast to the \citet{Kobay06} predictions. We also find the yields of Cr and Mn to be generally smaller and directly correlated with metallicity compared to the theoretical results. Our results for Ni are consistent with theory, although our model suggests that all Ni yields should be scaled up slightly. The outcome of this exercise is the computation of a set of integrated yields,
{\it i.e.}, stellar yields weighted by a slightly flattened time-independent Salpeter initial mass function and
integrated over stellar mass, for each of the above elements at several
metallicity points spanned by the broad range of observations. These results
are designed to be used as empirical constraints on future iron-peak
yield predictions by stellar evolution modelers. Special attention is paid to
the interesting behavior of [Cr/Co] with metallicity -- these two elements
have opposite slopes -- as well as the indirect correlation of [Ti/Fe] with
[Fe/H]. These particular trends, as well as those exhibited by the inferred
integrated yields of all iron-peak elements with metallicity, are discussed
in terms of both supernova nucleosynthesis and atomic physics. 

\end{abstract}


\clearpage

\section{INTRODUCTION}

The iron peak elements are synthesized in supernova environments.
The abundance data for these elements in Galactic halo and disk stars can 
provide important constraints on  the conditions 
({\it i.e.}, the elemental content of the ejecta, 
the supernova mass cut, explosive energies, etc.) 
that occur in explosive nucleosynthesis in, 
for example, Type II 
supernovae (SNe II). 
Abundance trends with metallicity for these elements,
{\it i.e.} galactic chemical evolution studies, 
can also provide insight into the individual contributions from 
both Type I (SNe Ia) and SNe II nucleosynthesis and 
into the progenitor masses for these objects.  
An examination of these iron-peak elements and the associated abundance trends, 
comparing stars with varying metallicities, also provides direct
insights into the chemical evolution of the Galaxy - 
quantifying the nature and 
frequency of the sources of the production of these elements.

These studies of iron-peak elements, 
unfortunately, have suffered due to a lack of precise 
abundance values. For example, it has been possible to obtain 
stellar photospheric abundances of Cr
from the moderately-strong (neutral) Cr I lines, which have accurately determined atomic 
properties, {\it i.e.}, log gf values. 
On the other hand, the weaker ionic Cr II lines commonly employed in abundance analyses 
tend to not have as well determined atomic properties (measurement associated errors; Sobeck et al. 
2007). There are significant discrepancies in the abundances 
determined using Cr I lines as opposed to Cr II lines (especially in metal deficient stars), which may be due to 
errors in the atomic properties of Cr II or in the atmospheric 
models and line transfer codes employed to determine these abundances (Sobeck et al. 2007). 

The abundance trends discussed in this paper were originally noticed 
by \citet{mw95}, who measured abundances of numerous elements in stars 
down to a metallicity of [Fe/H]=-4. In particular, they found that below 
[Fe/H]=-2.4 Cr/Fe decreased and Co/Fe increased with decreasing Fe/H. 
Thus, the Co/Cr ratio increases overall as Fe/H drops. They speculated 
that this particular behavior could be explained by the effects of 
metallicity or progenitor mass on stellar yields. In addition, alpha 
freeze-out, in which a high photon to baryon ratio in metal-poor stars 
results in an elevated volume density of alpha particles, favors the 
synthesis of nuclei heavier than Fe, e.g. Co, at the expense of 
lighter nuclei, e.g. Cr. At the same time, \citet{mw95} point out that 
the Mn/Fe ratio also increases with Fe/H and suggest that this may be 
linked to the contribution of Type~Ia supernovae, SNe Ia.

The extensive study by \citet{timmes95} employed detailed chemical evolution 
models and the massive star yields of \citet{ww95} in an attempt to model 
the evolution of numerous elements, including those of the Fe-peak, 
using extant data available at the time as constraints. Since the data did 
not extend below -3 in [Fe/H], the trends noted by \citet{mw95} were not 
noticeable in the plots presented in \citet{timmes95}. However, their models 
were successful in modeling the available data.

The trends in Cr/Fe, Co/Fe, and Mn/Fe were considered in light of supernova parameters by \citet{nak99}. Noting that stable Co and Fe are produced by isotopes originating in the complete silicon burning region during explosive nucleosynthesis, while Cr and Mn precursors come from the incomplete silicon burning zone farther out, these authors examined the effects of mass cut position, neutron excess, explosion energy, and progenitor mass on these element ratios. They found that the trends in the element ratios could be duplicated if the position of the mass cut migrated outward as metallicity increased. Varying the other three parameters independently appeared to have a less dramatic effect.

A major effort was made in this regard by \citet{francois04}, who compiled 
data for 12 elements, including six Fe-peak elements Ti, Cr, Mn, Fe, Co, 
and Ni, and analyzed the trends present in [X/Fe] vs. [Fe/H] plots. 
Their analysis centered on the computation of a detailed two-infall 
(timescale) 
chemical evolution model for the Milky Way. This particular model was 
originally developed by \citet{Chiappini97}, who used independent 
formation time scales for the halo-thick disk and the thin disk as 
well as a surface density threshold for star formation to reproduce 
numerous features of the Milky Way Galaxy. \citet{francois04} 
employed the massive star yields of \citet{ww95} for solar composition 
along with their own yield scaling factors to force matches to 
the data trends. Their final product is a table of recommended 
massive star yields derived empirically in this way, where the 
yield in solar masses is given as a function of stellar mass. 
A summary of their recommended changes to the original \citet{ww95} 
yields was also provided.

More recently, \citet{Kobay06} used new yield calculations to compute 
a chemical evolution model of the Galaxy. Their predictions for 
Fe-peak elements are generally in line with observations, although the 
specific behavior of Cr, Co, and Mn at low metallicity are not closely 
reproduced.

In summary, the trends found in the Fe-peak elements have inspired 
several promising theories about how they arise. At the same time, it is 
clear that the uncertainties involved in the computation of massive 
star yields are significant; the mass cut position and explosion 
energy for supernova models 
in particular would seem to be relatively unconstrained at this 
point. This situation makes it difficult to derive a reliable set of 
stellar yields.

The goal of this paper is to approach the problem from the reverse direction. By initially adopting the idea of \citet{francois04}, we employ detailed chemical evolution models to derive empirically a set of robust integrated yields, i.e. values arrived at by integrating stellar yields over an initial mass function. In this way we produce quantities which serve as real targets for future yield computations. That is, regardless of the assumptions going into the stellar yield computations regarding mass cut position or explosion energy, the theoretical predictions, when integrated over an IMF, must match the inferred integrated yields. The latter then serve as major constraints for yield studies for the Fe-peak elements studied here. As is explained below, this is largely possible because at [Fe/H] values below about -2, where the interesting trends appear, the instantaneous recycling approximation is a good one, and thus details of star formation history do not play a significant role. 
We also note that we have included many more data in our analysis 
than were available,
particularly at higher metallicities, for earlier studies, for example
those by   
Mc William (1997)\nocite{mw97}.

The sources for the observational data used in this study, along with the abundance trends, are discussed in section \ref{data}. Sections~\ref{analysis} and \ref{discussion} contain the results of our model studies of the observed trends. In section~\ref{summary} we summarize our work and list  our conclusions.

\section{DATA SOURCES AND ABUNDANCE TRENDS\label{data}}

For our pool of element abundance data, we draw from the following references: \citet{gratton91}, \citet{mw95}, \citet{fg98}, \citet{cayrel04}, 
\citet{sobeck06} and \citet{Lai08} \footnote{The \citet{sobeck06} analysis presents data only for Mn I.}.  In general, 
our sample consists of high-resolution spectral data and (combined) covers a
metallicity range of -4.0$\lesssim$[Fe/H]$\lesssim$+0.5.  With regard to exact element
abundances, there are data for these species: Ti I/II, V I/II, Cr I/II, Mn I, Fe I/II, Co I, and Ni I \footnote{Note that many of the ions do not have data
across the entire specified metallicity range and as a consequence, we focus primarily on abundances from the neutral species.}.  The specific 
observational characteristics associated with each of the
data sources are listed in Table \ref{sobeck}.  All of the chosen references employ the fundamental assumptions of plane-parallel geometry, one-dimensionality, 
and local thermodynamic equilibrium (LTE).  We also remark that each of the studies utilize different telescope-instrument set-ups, data reduction techniques, 
stellar model atmospheres, and line transfer codes (there is no single, consistent approach for our sample).

We point out that abundance analyses, which employ a three-dimensional or non-LTE methodology, provide a rigorous treatment 
of radiative transfer in stars.  However, such approaches up to this point have been limited in scope and have not yet been 
employed to determine abundances over the entire {\it observable range} of metallicity or in the many different
stellar populations (i.e. there is a limited amount of available data).  Furthermore, the magnitude of the change in the 
abundance data that occurs with the implementation of these methodologies depends 
upon a variety of factors including effective temperature, surface gravity, and metallicity (i.e. not all stellar abundances are affected {\it similarly}).  
Hence in order to assemble a large and cohesive data sample, we employ data exclusively from LTE analyses.  

Figure~\ref{6plot_notracks} displays the results of our data compilation for Ti I, V I, Cr I, Mn I, Co I, and Ni I, where in each panel
we plot [X/Fe] vs. [FeI/H]. Symbol color indicates the source of abundances, as defined in the legend.  Below we discuss each of the general element abundance trends
as a function of metallicity and remark upon any anomalous data points/behavior.

{\it Titanium}: In the upper left panel, the [Ti/Fe] abundance ratio appears to increase from its solar value to [Ti/Fe] $\approx$ +0.4 in the range
-1.0$\leq$[Fe/H]$\leq$+0.40.  It then hovers around [Ti/Fe]=+0.4 as metallicity is decreased with some scatter at the extremely metal-poor end
(especially from the \citet{mw95} data).  This trend in the Ti data is also seen in the abundances compilations from
\citet{timmes95} and \citet{francois04} over a similar metallicity range.

{\it Vanadium}: As shown in the upper middle panel, [V/Fe] remains roughly at its solar value across the full range of metallicity.  There are four aberrantly high
data points present from the \citet{mw95} sample at about [Fe/H]=-3.0.  We note that there are no vanadium abundances from \citet{cayrel04}.  A flat trend
is likewise displayed in the data plotted by \citet{timmes95}.

{\it Chromium}:  The upper left panel features Cr data.  For -2.1$\leq$[Fe/H]+0.4, Cr stays roughly constant with [Cr/Fe]$\approx$0.  Below [Fe/H]=-2.1, the [Cr/Fe]
ratio declines steadily with decreasing metallicity (again note these data are from Cr I transitions only).  A similar Cr behavior is
exhibited in the \citet{timmes95} and \citet{francois04} compilations.

{\it Manganese}: The Mn trend is displayed in the bottom left panel.  For the relatively metal-rich regime of -1.0$<$[Fe/H]$<$+0.4,
Mn decreases from a super-solar value to approximately [Mn/Fe]$\approx$-0.4.  It seems that the Mn abundance is flat in the 
region -2.5$<$[Fe/H]$<$-1.0 and then, there is a slight indication of a further downturn at [Fe/H]$\approx$-2.5 
(there is large scatter in the extremely metal-poor stars especially from the \citet{cayrel04} sample).  Results in both 
\citet{timmes95} and \citet{francois04} show a flat, {\it subsolar} behavior for [Mn/Fe] in the range -4.0$\leq$[Fe/H]$\leq$-1.0; above a metallicity
of [Fe/H]=-1, an upward [Mn/Fe] trend is evident.

{\it Cobalt}: Shown in the bottom center panel, Co behavior remains relatively flat at [Co/Fe]$\approx$0 in the range -2.2$\leq$[Fe/H]$\leq$+0.4.  
Then, the [Co/Fe] ratio proceeds to rise sharply as metallicity continues to decrease.  A duplicate Co trend is seen in \citet{francois04} 
while \citet{timmes95} lack Co data below [Fe/H]$\approx$-2.5.

{\it Nickel}:  As displayed in the bottom right panel, the [Ni/Fe] ratio hovers around its solar value for -2.0$\leq$[Fe/H]$\leq$+0.4.  For metallicities
lower than [Fe/H]=-2.0, the Ni behavior shows considerable scatter with little discernible trend.  This is in general agreement with what is exhibited
in both \citet{timmes95} and \citet{francois04}.

Two trends which are of particular interest to us are those of Cr and Co in the range -4.0$\leq$[Fe/H]$\leq$-2.2. The abundances of these
two elements seem to mirror one another in this metallicity regime-- see further discussion below in \S \ref{discussion}.

\section{COMPUTATION OF INTEGRATED YIELDS\label{analysis}}

We now use the trends shown in Fig.~\ref{6plot_notracks} along with 
chemical evolution models in order to derive the integrated yield 
$P_x$ as a function of metallicity for each of the six elements. We 
define an integrated yield of element {\it x} as
\begin{equation}
P_x \equiv   \int^{m_{up}}_{m_{down}} mp_x(m)\phi(m)dm\label{integratedyield}
\end{equation}
where $p_x(m)$ is the stellar yield, $\phi(m)$ is the initial mass 
function, and $m_{up}$ and $m_{down}$, are, respectively, the upper and 
lower limits to the mass range of all stars formed. $P_x$ is then the 
mass fraction of all stars formed within the mass range that is eventually 
expelled as new element $x$. 

The benefit of integrated yields is that they provide a convenient means of directly comparing different sets of published stellar yield predictions, where normally those yields are presented as a function of stellar mass. The stellar yield as a function of mass may vary widely, depending upon the assumptions made regarding mass cut, explosion energy, or other free stellar parameters. But in the end, a set of yields when integrated over a mass function, i.e., the integrated yield, must provide an adequate amount of each element to explain observed abundance patterns, regardless of the assumptions which went into the original yield calculations.

To derive integrated yields from the observed trends discussed above, we 
employed a chemical evolution code to compute a track through each [X/Fe] 
vs. [Fe/H] plot, adjusting the yields by scaling factors until we 
achieved suitable agreement between observation and theory. At that point 
the yields for each element were integrated over an IMF to produce a 
value for the integrated yield.

Our chemical evolution code is the one used most recently by \citet{HP07} and is described in detail in the appendix of that paper. Briefly, the code carries out a time integration for a single zone characterized by a star formation history specified by an infall rate, an IMF which was a slightly flattened ($\alpha$=-1.20) Salpeter relation \citep{Salpeter55}\footnote{We found that an IMF slope of -1.20 resulted in a better fit to the observed age-metallicity relation. This flattened Salpeter IMF is roughly consistent with the slope required ($\alpha$=-1.10) to reproduce some of the N/$\alpha$ versus $\alpha$/H data reported in \citet{p02}.}, a star formation efficiency, and a star formation law. At each time step the new production of each element, obtained by integrating the stellar yield as a function of mass over the effective mass range, is added to the present level in order to update the current abundance level of that element in the interstellar medium. Thus, the program keeps track of the abundance of each element as a function of time and metallicity. Generally speaking, the relative contribution of stars of a particular mass is directly linked to the rate of star formation at the time in history when these stars were formed. The elements included in the calculation were H, He, C, N, O, Ne, Si, S, Ti, V, Cr, Mn, Fe, Co, and Ni. The values for the basic set of input parameters used in all models, unless otherwise noted, are provided in Table~\ref{parameters}. Many of these values were adopted directly from \citet{timmes95} and our models were then checked for consistency against their results. We were able to reproduce the metallicity distribution function in their Fig.~38, the age-metallicity relationship in their Fig.~7, as well as their tracks in the  [X/Fe] versus [Fe/H] plots relevant to the elements which we are investigating in this paper (adopting their yields).

We began the modeling by using the combination of massive star yields by \citet{Portinari98} for H through S and \citet{Kobay06} for Ti through Ni (see the list above). For low and intermediate mass stars we adopted the yields of He, C, and N by \citet{Marigo01}. To determine the contributions of Type~Ia supernovae we employed the prescriptions of \citet{mg86} along with the yields published by  \citet{nomoto97}. For the massive stars, Kobayashi et al. publish yields for both Type~II supernovae and hypernovae. We assumed, as they did, that these two event types occur with equal frequency, and thus the stellar yield at a particular mass was set equal to the average of the supernova and hypernova yield\footnote{Note that the Fe-peak yields of \citet{Kobay06} have been piggy-backed onto our basic code which has long used the stellar yields of \citet{Portinari98} and \citet{Marigo01} in tandem, as these two sets were produced by the same study and are therefore designed to be consistent over their effective mass range. While adopting the Kobayashi yields for elements He-S may at first seem preferable, in these calculations it is only the metallicity which matters in determining the value of the stellar yields at any point in time.}. Note that the choice of the Kobayashi yields was arbitrary, as they only provided a starting point in our search for an empirical set of yields. Our outcome is not based upon our choice of the initial set of trial yields.

Figure~\ref{6plot_tracks} shows the observed abundances in 
Fig.~\ref{6plot_notracks} but now with model tracks added. The 
solid green lines show the model results using the Kobayashi 
yields as published and unscaled. One can see that in the cases 
of Ti/Fe, V/Fe, Co/Fe, and perhaps Ni/Fe  the model underpredicts the 
observed abundances, particularly at low metallicity. On the other hand, 
the same model predicts Cr/Fe levels above those observed but does well 
with Mn/Fe. In the case of V/Fe, the data are ambiguous, as it is 
unclear whether the ratio increases below [Fe]=-2, as indicated by a 
few of the \citet{mw95} points, 
or remains constant, as suggested by some data points from McWilliam et al. and \citet{gratton91}. 

In an alternative version of this model we employed the scheme for calculating the contributions of SNIa explosions discussed by \citet{Matteucci06}. This method, based upon empirical evidence, assumes the existence of a bimodal distribution of delay times, wherein 35-50\% of the Type Ia progenitors have lifetimes around 10$^8$ years, while the remaining systems involve small mass progenitors which require more time to evolve and have a broad distribution of delay times. This bimodal feature results in a significant number of Type~Ia events beginning to occur at metallicities as low as [Fe/H]=-2.0, i.e., significantly earlier than the prescription of \citet{mg86} would predict. The result of using the \citet{Matteucci06} scheme is shown with the dashed green line. Clearly there is no general improvement in the match between data and theory. 

We note the significant scatter at low metallicity for most of the elements in Fig.~\ref{6plot_tracks}, which makes interpretation of the trends more difficult. The possible explanations for this scatter include the lack of sufficient data, the presence of the ejecta from individual nucleosynthetic events such as the first stars, and insufficient mixing of that ejecta with the interstellar medium. 

We next calculated a model using the massive star yields for Ti-Ni by \citet{ww95}, who listed isotope yields prior to decay. [Recall that \citet{francois04} used these yields in their analysis.] The isotopes whose individual yields contribute to the final yield for a particular Fe-peak element, either directly or through decay, are listed in Table~\ref{isotopes}. The solid red line shows the results in Fig.~\ref{6plot_tracks}, and we point out that our tracks agree closely with the model results of \citet{timmes95}, as expected. The quality of the match between observation and theory varies from element to element. For Ti/Fe, our computed track closely follows the one employing the \citet{Kobay06} yields but neither reproduces the upward trend with declining metallicity. Roughly the same can be said for the case of V/Fe. For Cr/Fe our model reproduces the flat behavior of the data at metallicities above -2, but below that point the model has less of a downturn than the data. The model produces a good match with the data in the case of Mn/Fe, but falls well below the Co/Fe trends. Finally, in the case of Ni/Fe the WW95 model is consistent with the data at both the high and low metallicity ends, while sagging below the observations in the mid-range.

In order to improve the model fit to the observations and thus allow us to compute useful integrated yields, we applied scaling factors to the Kobayashi yields, thereby forcing a fit to the data. These factors were derived through trial and error and are listed in Table~\ref{scalingfactors}. In particular, for the Fe peak elements other than Fe itself\footnote{We used unscaled Kobayashi yields for Fe over the entire metallicity range. This was deemed reasonable, since their chemical evolution model of the solar neighborhood utilizing these yields ably reproduced both the observed age-metallicity relation as well as the metallicity distribution function (see their Fig.~6b,c)}, we adjusted the massive star yields in a succession of trial models until we obtained a reasonable eyeball fit to the data over the metallicity range from -4 $<$ [Z] $<$ +0.5.   The model results in this case are shown in Fig.~\ref{6plot_tracks} with a solid violet line. We see that in the cases of Ti, V, Mn, and Co the trends are matched fairly well. We note that for V,  a few of the points from the data of \citet{mw95} suggest an upward trend below a metallicity of -2. Thus, we used a second scaling factor to produce a model in which the behavior of those points in particular was reproduced. In the case of Ni, the scatter in the data below -2 is too large to speculate about the success of the model. Finally, the situation with Cr will be discussed further below as it relates to Co, but we simply point out here that while the apparent plateau below -3 is unexplained by our  force fit model, the calculations employing either the \citet{Kobay06} or \citet{ww95} yields do produce a flattening in that low metallicity region.

We next derived the values for massive star integrated yields of the 
Fe-peak elements based upon our force fit model just described. We did this 
by rerunning the model incorporating the scaled Kobayashi yields and at each time (metallicity) point integrated the scaled yields 
over the same IMF as that used in the model calculations, i.e. a slightly flattened ($\alpha$=-1.20) Salpeter IMF, according to equation~\ref{integratedyield}. The 
integrated yields derived in this manner are provided in 
Table~\ref{integratedyields}.\footnote{A crucial point to make here is 
that the Fe-peak elements at low metallicity (early times) are 
principally forged by massive stars with short lifetimes, i.e., a few 
million years, relative to chemical evolution time scales, i.e., a few 
billion years. Thus, the familiar instantaneous recycling approximation 
applies and it becomes unnecessary to account for the small differences 
in stellar lifetimes over the mass range of massive stars. Were this 
not true, then the details of star formation history 
would begin to play a role when comparing the yields of two elements 
and the integrated yields would not be as meaningful.} 
Note that the integrated yields for V correspond to the upper model
track.

The first two columns of the table give the log of the metallicity and 
its value normalized to the sun \citep{AGS05}, respectively, while subsequent columns 
list the log of the individual integrated yields. We emphasize that these 
integrated yields provide useful constraints for assessing published 
stellar yields, since any set of yields when integrated over an IMF 
should closely match them. Note that the IMF may have any form as long
as the integral of the yields over that IMF gives the proper value for the integrated yield that is consistent with the observations. 

The integrated scaled Kobayashi yields are 
plotted in Fig.~\ref{P7} as a function of the log of metallicity 
(violet tracks) along with analogous values for the unscaled yields 
from the same authors (green tracks). (Since we did not use 
scaled Fe yields, only the unscaled integrated yield track is plotted for that element.) 
Clearly the largest offsets occur at metallicities below -2. In the cases of Ti, V, and Co, we find a clear decrease in the integrated yields as metallicity rises. This contrasts with the predictions of \citet{Kobay06}, who find relatively little metallicity sensitivity for the yields of these three elements. On the other hand, for Cr and Mn we infer a direct relation with metallicity for the yields of these elements, while again \citet{Kobay06} find little metallicity effect. For Ni, our results are qualitatively similar to those of  \citet{Kobay06} below a metallicity of -1, after which we find that Ni yields decline while they infer a rise with increased metallicity. Quantitatively, the theoretical yields fall below our values at metallicities below solar.

While our general approach to yield evaluations resembles that of \citet{francois04}, the details of the studies are different enough that a direct comparison of results is difficult if not impossible to make. For example, they scaled the yields of \citet{ww95} for solar metallicity only, while we employed the \citet{Kobay06} yields and accounted for metallicity effects. (Note that the latter set of yields was unavailable at the time of the Fran\c{c}ois et al. study.) Obviously, a valid comparison would be possible only if the two research teams had based their studies on the same set of theoretical yields. We do point out, however, that both groups were successful at producing a model with scaled yields that resulted in good fits to the observations.

Finally, we compare our integrated yields from Table~\ref{integratedyields}, of all seven of the Fe-peak elements we are considering, in Fig.~\ref{Px} as a function of metallicity log~Z. Here we see some interesting trends which we will attempt to interpret in the next section. First, Cr production increases while Co production decreases as a function of metallicity, resulting in the two tracks crossing at around log~Z/Z$_{\odot}$ of -3. This result clearly is linked directly to the observed behavior of these two elements with metallicity (see data for these two elements in Fig.~\ref{6plot_notracks}). Other interesting trends include the indirect behavior of V and Ti with metallicity.

Under conditions such as we have at low metallicity where the instantaneous recycling approximation applies, one can show using standard chemical evolution equations \citep{matteucci01} that the only factors besides the integrated yields which might substantially influence the evolution of an element ratio are gas infall or outflow, through their diluting/concentrating effects. Therefore, we ran two variations of our force fit model in which we employed  star formation time scales of 2~Gyr and 7~Gyr, where we previously used 4~Gyr in the basic model (see Table~\ref{parameters}). We found no perceptible difference between the three models, suggesting that our integrated yield values are robust within a wide range of galaxy formation conditions and regions.

\section{DISCUSSION\label{discussion}}

\subsection{Abundance Ratios and Metallicity Trends\label{ratios}}

As discussed in the Introduction, the particular behavior of Cr/Co with increasing metallicity was 
originally pointed out by \citet{mw95} and deserves special attention. 
This behavior is illustrated in Fig.~\ref{cr2covfel}, where to the data displayed in Figs.~\ref{6plot_notracks} and \ref{6plot_tracks} we have added abundances from \citet{Barklem05} indicated with green diamonds. Also shown in the figure are five well-studied
r-process-rich halo stars, indicated by red circles. 
The abundance ratios of Cr/Co in all of these high-resolution,  high 
S/N studies (\cstwo, [Fe/H] = --3.1, 
Sneden \etal\ (2003)\nocite{sne03};
\hdone, [Fe/H] = --3.0, 
Westin \etal\ (2000)\nocite{wes00}; HD 122563, [Fe/H] = --2.7, 
Westin \etal\ (2000)\nocite{wes00}; \bd, [Fe/H] = --2.1, 
Cowan \etal\ (2002)\nocite{cow02}; \hdtwo,
[Fe/H] = --2.2, Ivans \etal\ (2006)\nocite{iva06}) 
are consistent with  the behavior of the other sample stars
illustrated here.

The data together show a steady 
increase by a factor of more than 10 in Cr/Co between -4 and -1.5 of [Fe/H], 
followed by a leveling off at roughly -1.5 as metallicity continues
to increase. For comparison, data presented in \citet{timmes95}, which extend down only to -3 in metallicity, show a slight decline in Cr/Fe, while Co/Fe remains level at their lowest metallicities, although the scatter is enough that 
it is difficult to discern a downward trend in Cr/Co [\citet{timmes95} do not explicitly plot Cr/Co].

In Fig.~\ref{cr2covfel} we also show the predictions of several 
chemical evolution models, where the line types and colors are consistent with those used in Fig.~\ref{6plot_tracks}. Here we see that the model which employed the scaled Kobayashi yields and successfully matched the 
Cr and Co trends with Fe in Fig.~\ref{6plot_tracks} satisfactorily reproduces 
the trend in Cr/Co in Fig.~\ref{cr2covfel} (solid violet track), as expected. This is true at least out to -3.5 in metallicity, below which the data suggest the presence of a plateau, which is not predicted by our model.  
The model using the Kobayashi yields as published (solid green track), along with two variations of it in which all core collapse events are either SNII (solid maroon track) or hypernovae (solid orange track), are generally consistent with the data above a metallicity of -2 but fail to reproduce the downward trend at lower metallicities.  Also, models in which the SNIa scheme of \citet{Matteucci06} was used are indicated with dashed lines of the same color as the associated model that makes use of the same stellar yields and input parameters. Clearly, this change does not produce an improvement to the original models. Finally, the solid red line represents 
the model in which we employed the yields and input parameters of 
\citet{ww95} and \citet{timmes95}, respectively.  This track does not match the data well except near solar metallicities, due in large part to the predicted behavior of Co/Fe (see Fig.~\ref{6plot_tracks}).

In general, the yields of \citet{Kobay06} coupled with either an equal mix of SNII and hypernovae or 100\% hypernovae give reasonable matches to the data at metallicities above -1.5. At metallicities below this, however, the nearly level predicted values of Cr/Co for these two models fail to match the downward trend. Also, the all-SNII model predicts Cr/Co values which are generally too high at all metallicities.

An additional comment concerns the apparent plateau below about -3.5 in metallicity. Clearly, our forced fit using scaled Kobayashi yields does not reproduce this behavior. At the same time, however, we note that the model using the WW95 yields does predict a roughly constant value for [Cr/Co] over this metallicity range. Combining the WW95 results below -3 with the scaled Kobayashi model results above that level would clearly produce a good fit to the data.

Finally, we see that the general upward trend of [Cr/Co] with metallicity for [Fe/H] below -1.5 in Fig.~\ref{cr2covfel} is explained by increasing Cr integrated yields and decreasing Co integrated yields below [Z]=-1.2 in Fig.~\ref{Px}. Above that same point, while Cr yields appear to level off, the Co yields actually reverse direction and go up. According to the discussion in \citet{nak99}, the behavior of Cr and Co yields below -1.2 may be linked to a systematic outward migration of the mass cut with metallicity during the SN explosion.

The same models featured in the above analysis of [Cr/Co] are shown in 
Fig.~\ref{ti2fevfe}, which displays the data and models relevant to [Ti/Fe]. 
Again we 
see that the scaled Kobayashi model successfully reproduces the trend in 
the data, while the others are less successful. In particular, the model 
employing the \cite{ww95} yields (solid red track) fails by two orders of 
magnitude to reproduce the [Ti/Fe] behavior. Referring again to Fig.~\ref{Px} 
we see that the observed downward trend in [Ti/Fe] is consistent with a 
similar trend in the integrated yield of Ti. 

\subsection{Recent Findings for Cr and Co}

There have been two recent laboratory determinations of atomic data (e.g. oscillator strengths) for both the neutral (Sobeck et al. 2007\nocite{sobeck07})
and first-ionized species (Nilsson et al. 2006\nocite{Nilsson06}) of chromium.  The study by Lai et al. (2008\nocite{Lai08}) was the only one of our 
selected sample to employ these new data.  They found that [CrI/Fe] ratio decreased in the region -4.3 $\leq$ [Fe/H] $\leq$ -2.0 
while the [CrII/Fe] ratio remained solar over the same metallicity range.  As a follow-up, they generated plots of Cr I and Cr II as a function of effective 
temperature (T$_{eff}$; a stellar atmospheric parameter) and discovered a trend in the Cr I data.  Their finding may indicate that the neutral
Cr abundances are affected by departures from LTE.

Similarly, there have been up-to-date laboratory determinations of oscillator strengths for both the neutral (Nitz et al. 1999\nocite{Nitz99}) and first-ionized
species (Crespo Lopez Urratia et al. 1994) of cobalt. The abundance investigation of 17 stars by Bergemann (2008\nocite{Bergemann08}), 
which employed a non-LTE methodology and these new atomic data, found that the [CoI/Fe] ratio increases as metallicity decreases in the 
range -2.5 $\leq$ [Fe/H] $\leq$ 0.0.  This result contradicts the data from our selected studies and signifies that Co I abundances are
susceptible to non-LTE effects.  It is evident that further examination of both the Cr and Co abundance trends with [Fe/H] over 
the full extent of metallicity and in a statistically-significant data sample is warranted to validate these two recent findings.

\subsection{Supernova Nucleosynthesis Explanations}

As discussed above and indicated by red, filled circles 
in 
Fig.~\ref{cr2covfel},
we included five  well-studied, metal-poor r-process-rich
stars in our abundance comparisons of Cr/Co.
Since the suggested site for the r-process is core-collapse 
(massive star) supernovae 
(see {\it e.g.}, Sneden, Cowan, \& Gallino 2008\nocite{sne08}),  
the products of (early Galactic) 
nucleosynthesis in these halo stars, including the 
iron-peak elements, have not had major contributions from 
Type Ia SNe -   those events  presumably arising from objects that
require much longer stellar evolutionary timescales than Type II SNe.
There have been several attempts to explain the behavior of 
Cr/Co versus metallicity based upon Type II supernova nucleosynthesis.
Thus, for example it has been shown that Co and Fe result from complete
Si burning, while Cr is synthesized in these models from incomplete
Si burning (Nakamura {\it et al.} 1999\nocite{nak99}). 
We see in Fig.~\ref{6plot_notracks} that [Co/Fe] and [Cr/Fe] have opposite
slopes with respect to metallicity. 

We note that the results of Nakamura et al. supernova models
depend critically upon
the mass cuts between the nascent neutron star (or black hole)
and the ejected
envelope. The yields of the synthesized material depends upon the 
ejected mass, and thus directly upon these mass cuts. 
How these mass cuts depend upon progenitor masses and upon
the explosion energies is also critical to determining 
the elemental yields.
In addition there may be other contributing factors in these models
such as rotation energy that could  affect the nucleosynthesis 
products.

\subsection{Ionization States, Atomic Data and Abundances}

The selected abundance investigations do have a few weak points.  For instance, they do not employ the most recent determinations of atomic data (with exception of V I, 
V II, and Fe I).  Additionally, there is a general presumption of iron as a reliable and robust abundance indicator.  However recently, Lai et al. (2008\nocite{Lai08}) 
found a correlation between excitation potential and both metallicity and effective temperature for Fe I transitions (in a sample of extremely metal deficient stars).  
As an alternative, \citet{ki04} suggest the exclusive employment of Fe II transitions to derive metallicity (and thereby, avoid the issues associated
with the use of Fe I lines).  None of the chosen investigations did this in their respective determinations of metallicity.   

Also for the majority of stars examined (with low temperature and metallicity), the dominant form of all of the 
elements is the first-ionized species (especially the elements with an ionization potential less than 7 eV: Ti, V, Cr).  Yet, the selected studies largely rely upon 
transitions only from the neutral species of the element (and in fact, do not present any data for Mn II or Co II).

Then, the derivation of abundances from Fe-peak elements is sensitive to departures from LTE (effects such as overionization and resonance
scattering; e.g. Shchukina \& Trujillo Bueno 2001\nocite{Shch01}).  For example, the abundance investigation of 14 stars by Bergemann \& Gehren (2008\nocite{Berg08}), which 
employed a non-LTE methodology, found that the [MnI/Fe] ratio remained solar in a metallicity range -2.5 $\leq$ [Fe/H] $\leq$ 0.0.  This finding contradicts 
the results of the selected abundance studies.  Note though, that the magnitude of these non-LTE effects (for each of the elements) has yet to be quantified in 
rigorous and consistent analysis of a large stellar sample.

\section{SUMMARY \& CONCLUSIONS\label{summary}}

The principal goal of this paper has been to produce a set of integrated stellar yields which can be used by theorists in the future to test their yield predictions for Fe-peak elements.
Utilizing an extensive data base of published stellar iron peak elemental
abundances for the solar vicinity, we have produced plots of the form [X/Fe] versus [Fe/H] for the elements Ti, V, Cr, Mn, Co, and Ni. We next employed detailed one-zone chemical evolution models to evaluate the massive star yields of \citet{ww95} and \citet{Kobay06} in their ability to reproduce these trends. Finally, we scaled the latter yield set and used it as input to our models in order to force a fit to the data for each plot. From this scaled set we then derived our empirical integrated yields, using a slightly flattened time-independent Salpeter initial mass function. Our analysis led to the following conclusions:

\begin{itemize}

\item As recognized previously by other authors, there are clear upward trends of [Ti/Fe] and [Co/Fe] and downward trends of [Co/Fe] and [Mn/Fe] as metallicity decreases. This is especially the case below [Fe/H] of roughly -2.

\item The above trends appear to be the result of changes in massive star yields with metallicity and are unrelated to star formation history.

\item Models utilizing yields of \citet{Kobay06} and \citet{ww95} generally reproduce the data near solar metallicity. However, the agreement between observation and theory at lower metallicities is generally less satisfactory.

\item A set of integrated yields as a function of metallicity was derived from global data trends for six elements.

\end{itemize}

In applying scaling factors to the yields of \citet{Kobay06} to force a fit to the observations, we make the following comparisons between the empirical and theoretical yields. Our results suggest that actual yields of Ti, V, and Co are generally larger as well as  anticorrelated with metallicity compared to the \citet{Kobay06} predictions. We also find the yields of Cr and Mn to be generally smaller and directly correlated with metallicity compared to the theoretical results. Our results for Ni are consistent with theory, although our model suggests that all Ni yields should be scaled up slightly.

\begin{figure}
   \centering
   \includegraphics[width=6in,angle=270]{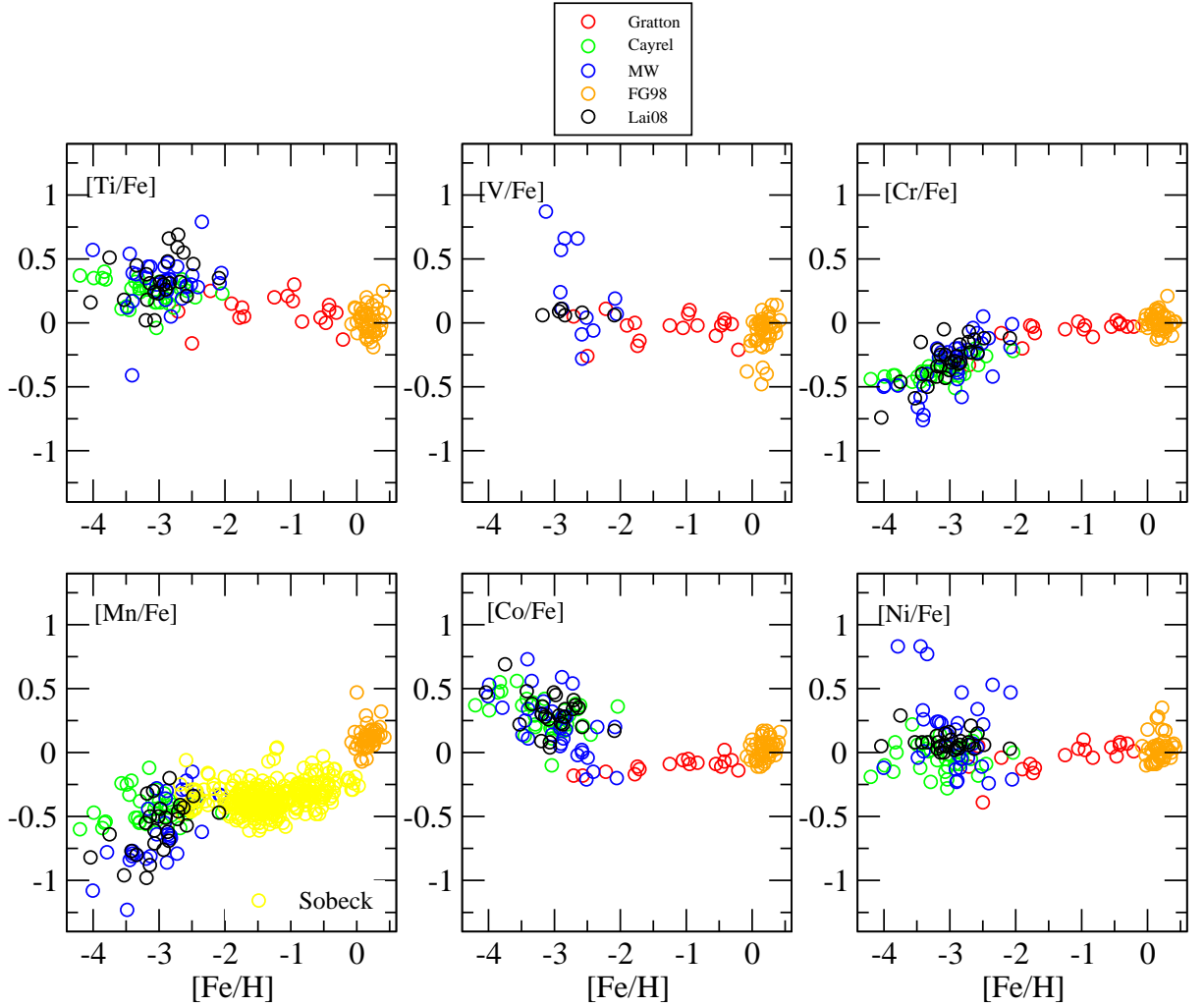} 
   \caption{Plots of [X/Fe] versus [Fe/H], showing observed values published in those sources indicated by symbol color defined in the legend. The specific ratio representing the dependent variable is indicated in the upper left corner of each panel.}
   \label{6plot_notracks}
\end{figure}

\begin{figure}
   \centering
   \includegraphics[width=6in,angle=270]{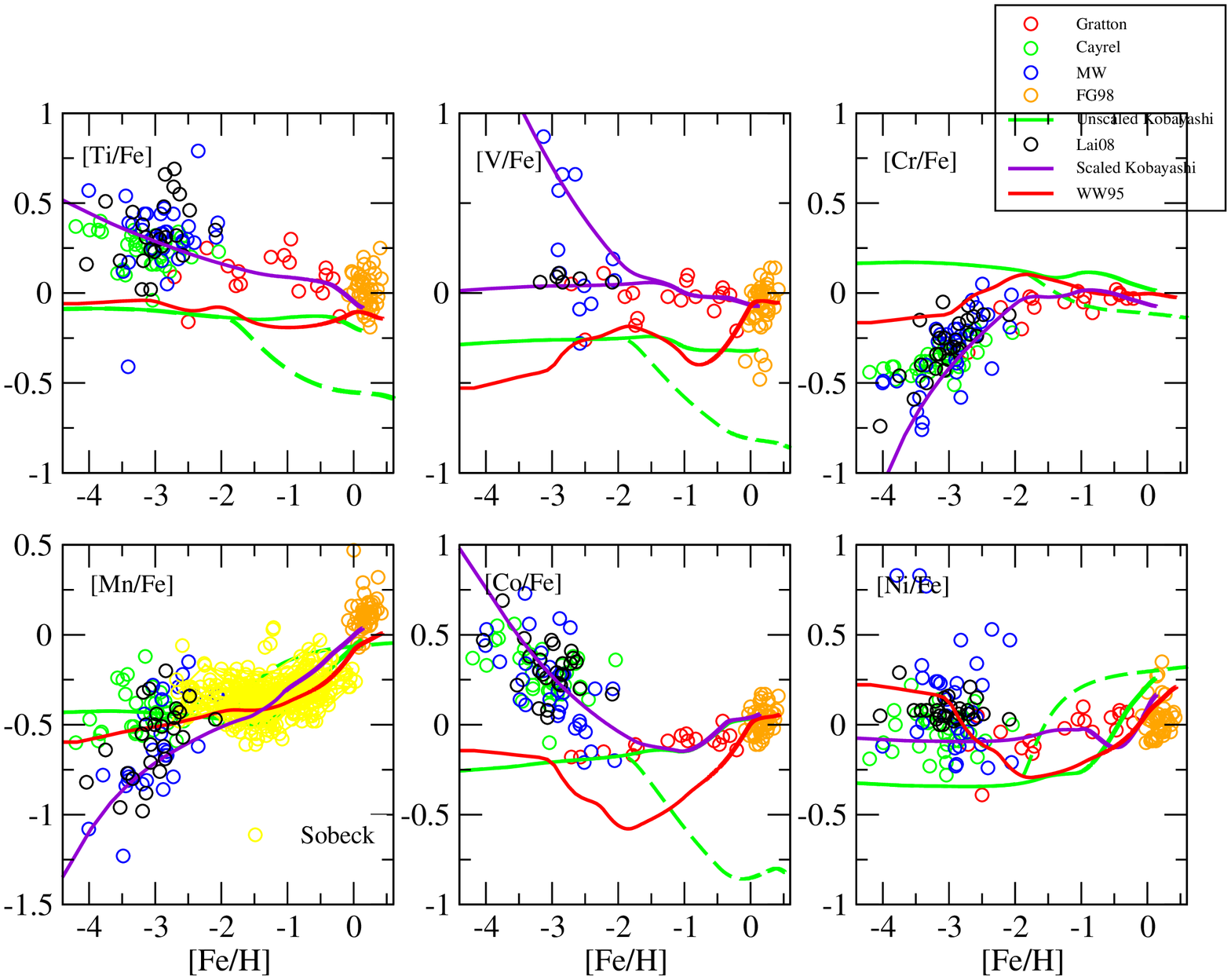} 
   \caption{Like Fig.~\ref{6plot_notracks} but with model results included. The green tracks are from models which employ the original Fe-peak yields of \citet{Kobay06}, while the violet tracks are from models in which the scaled Kobayashi yields have been used. The red tracks refer to a model that employs the massive star Fe-peak yields of \citet{ww95}. The green dashed tracks show the results when details of the prescription of \citet{Matteucci06} for SNIa events is employed. The legend shows the symbol and color for each data set of observed abundances. }
   \label{6plot_tracks}
\end{figure}

\begin{figure}
   \centering
   \includegraphics[width=6in,angle=270]{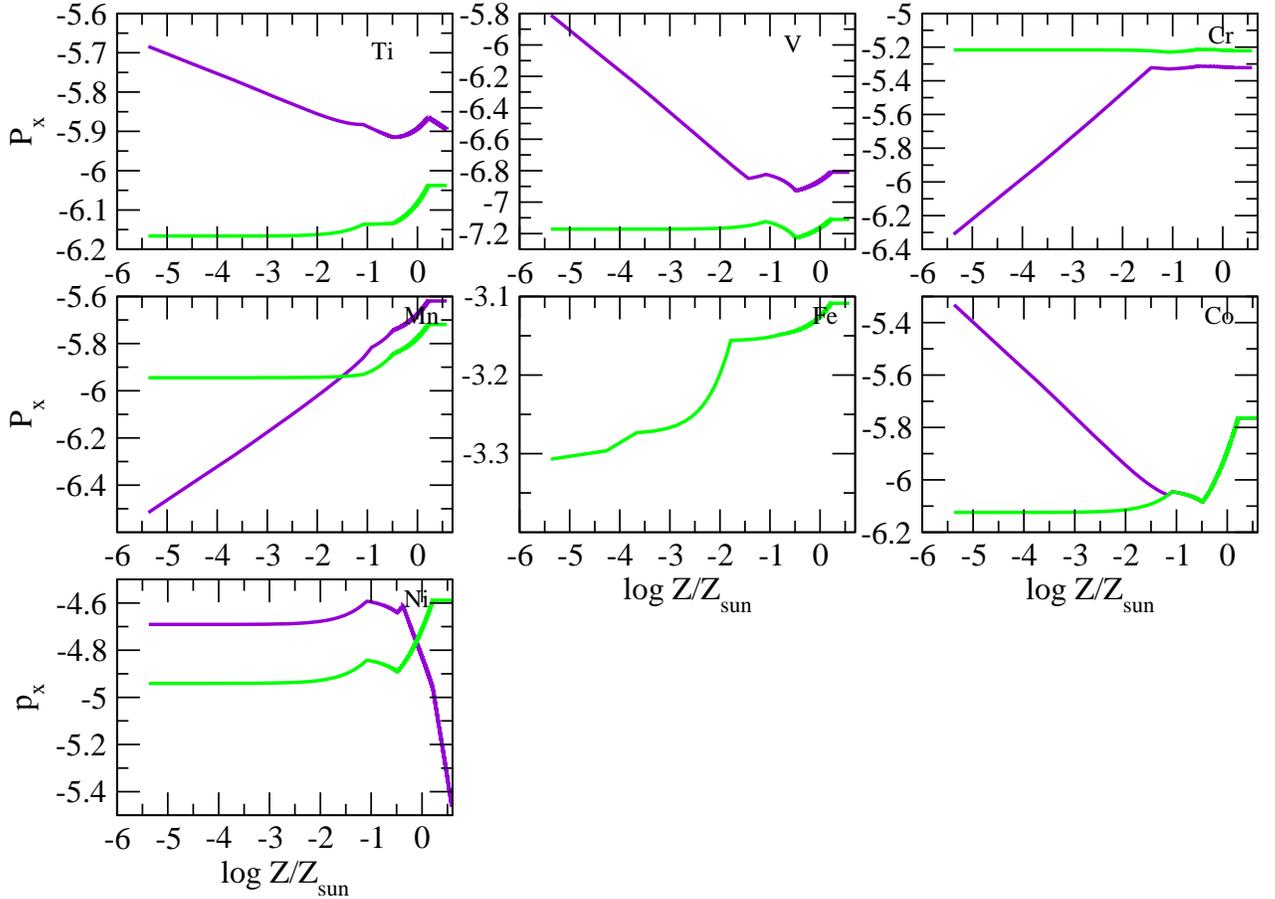} 
   \caption{The integrated yield P$_x$ versus log Z/Z$_{\odot}$ for the seven elements of the Fe peak under consideration. Green tracks refer to unadjusted yields from \citet{Kobay06}, while violet tracks show values for the adjusted yields. The value for Z$_{\odot}$ was assumed to be 0.0122 [\cite{AGS05}.]}
   \label{P7}
\end{figure}

\begin{figure}
   \centering
   \includegraphics[width=6in,angle=270]{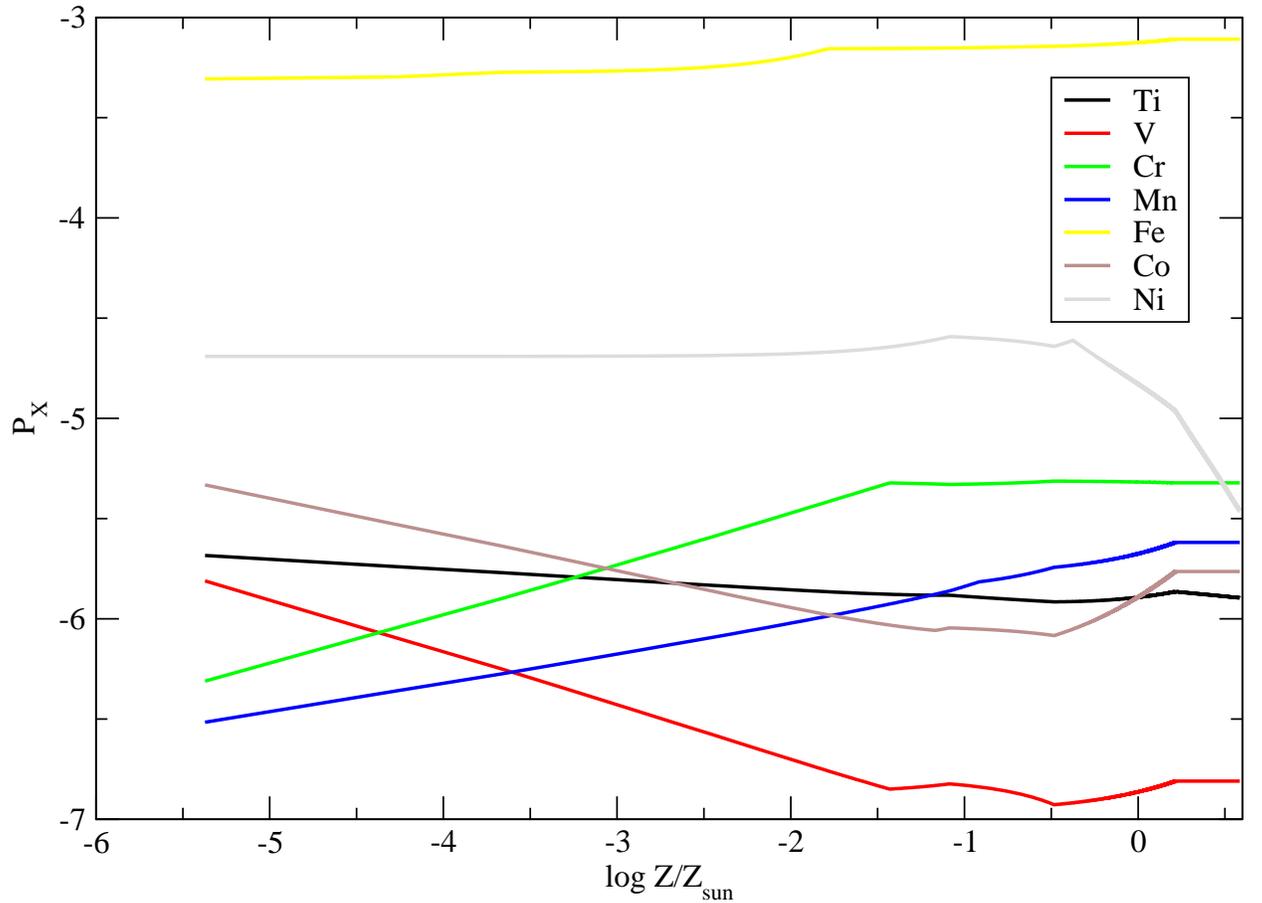} 
   \caption{A comparison of integrated yields derived from the adjusted Kobayashi yields (the violet tracks in Fig.~\ref{P7}) for each of the seven Fe peak elements as a function of log Z/Z$_{\odot}$. The legend connects line color with the element. The value for Z$_{\odot}$ was assumed to be 0.0122 [\cite{AGS05}.]}
   \label{Px}
\end{figure}

\begin{figure}
   \centering
   \includegraphics[width=6in,angle=270]{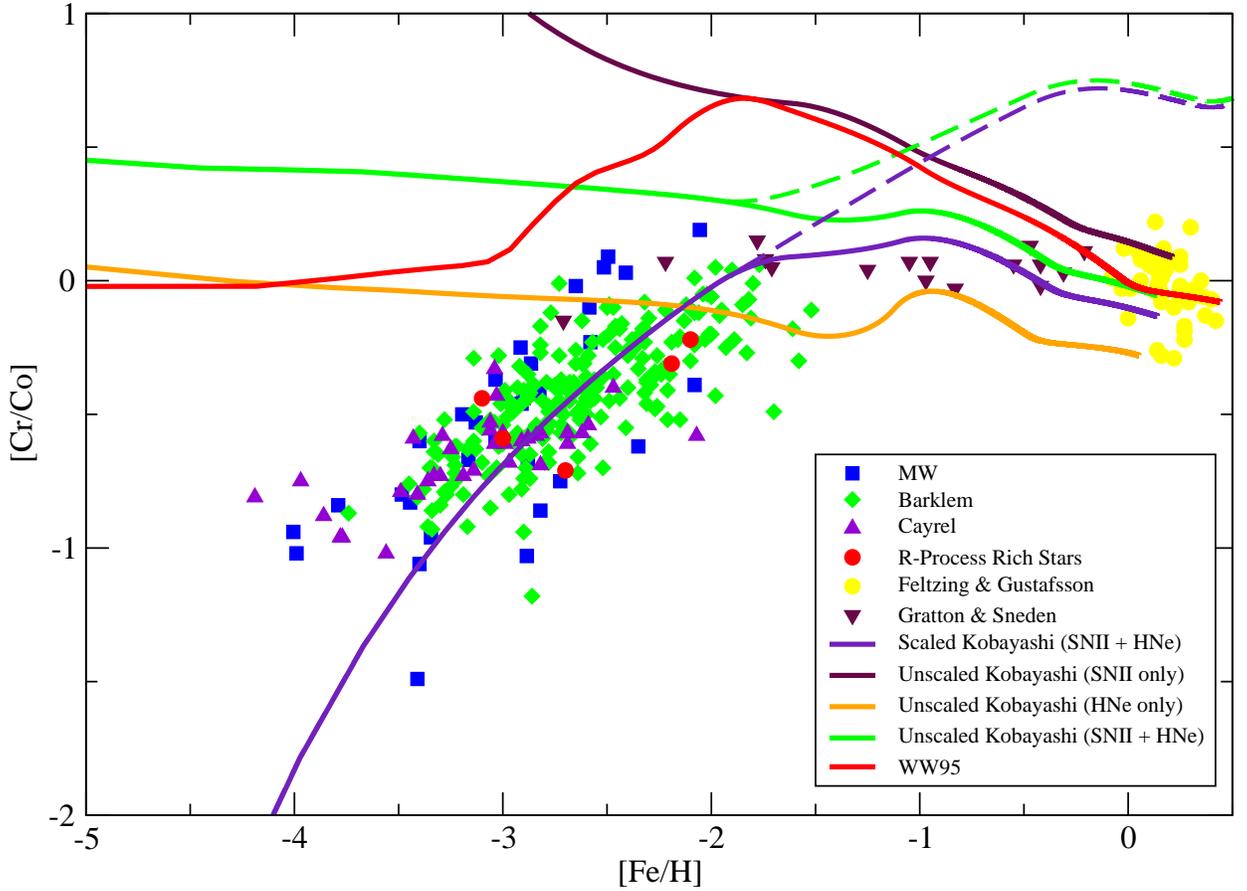} 
   \caption{[Cr/Co] versus [Fe/H]. 
Data plotted
are neutral species of the elements.
Data sources are distinguished by 
symbol shape and color and are identified in the legend. 
Curves show the results of several chemical evolution models, 
also identified in the legend and explained in the text. The dashed green and violet lines show the effect of using the SNIa prescription of \citet{Matteucci06} instead of that of \citet{mg86}.}
   \label{cr2covfel}
\end{figure}

\begin{figure}
   \centering
   \includegraphics[width=6in,angle=270]{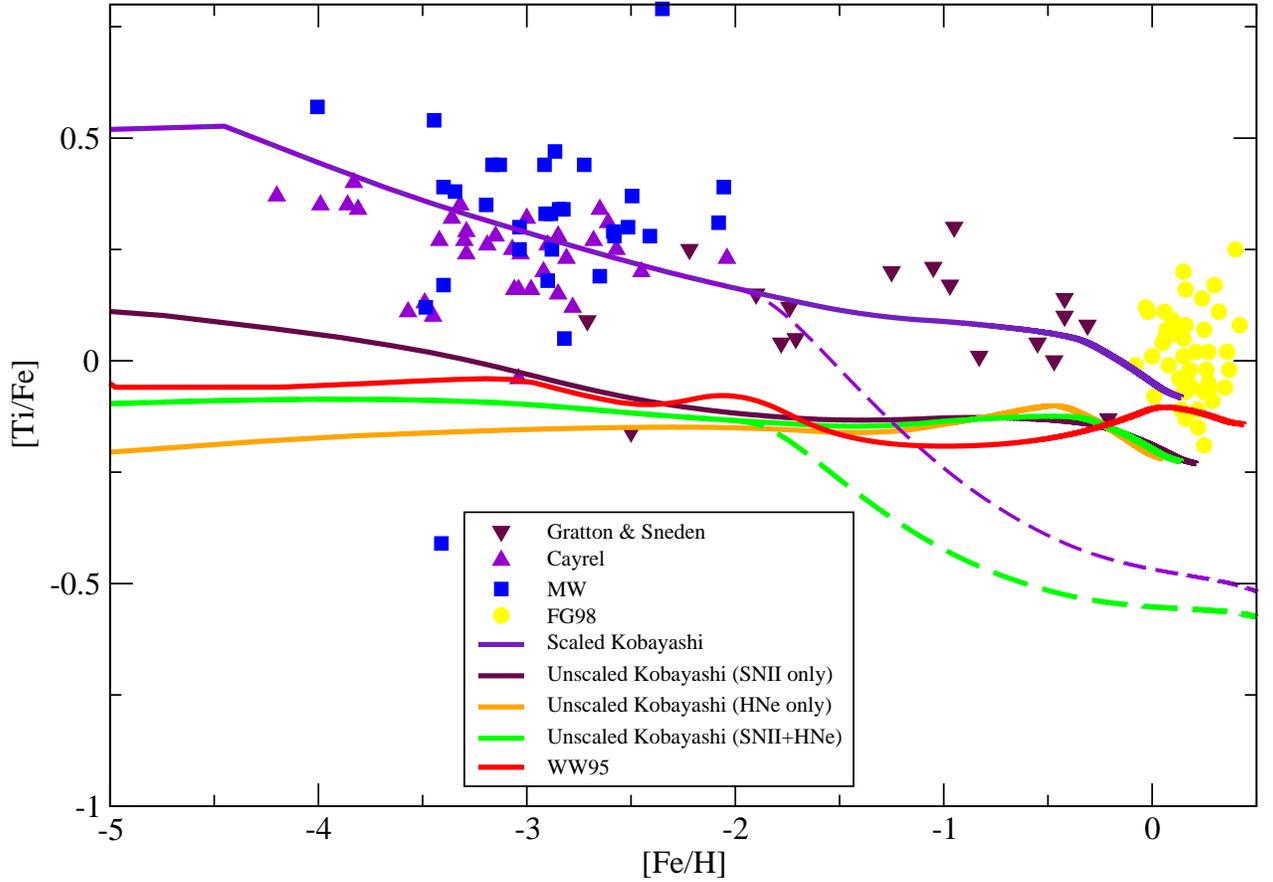} 
   \caption{Like Fig.~\ref{cr2covfel} but for [Ti/Fe] versus [Fe/H]. Data sources are distinguished by symbol shape and color and are identified in the legend. Curves show the results of several chemical evolution models, also identified in the legend and explained in the text. The dashed green and violet lines show the effect of using the SNIa prescription of \citet{Matteucci06} instead of that of \citet{mg86}.}
   \label{ti2fevfe}
\end{figure}

One clear problem with our analysis involves the scatter in the data and in some cases the difficulty in establishing the nature or direction of the trend. A good example can be seen in the plot of [V/Fe] versus [Fe/H] in Fig.~\ref{6plot_notracks}, where we see both a possible flat as well as an upward trend below [Fe/H]=-2. A second example appears in the graph of [Cr/Fe] versus [Fe/H] at metallicities below -3. Here we see an apparent bifurcation in the data as one branch remains horizontal while the other continues trending downward. In the case of V, we have attempted to match both branches, while in the Cr case we ignore the horizontal branch for now. At these low metallicities it is likely that the abundance pattern observed in a star is the result of expelled material from only one earlier-generation star instead of a well-mixed contribution from many such stars. Thus, star to star variations echo analogous differences in the yields of the earlier stars. This is not an effect which our models are designed to take into account. Rather homogeneous mixing is assumed. Future work should attempt to explore the causes of the observed scatter and its possible link to apparent bifurcations in the data. For now we have attempted only to analyze what we see as global trends in each plot in Fig.~\ref{6plot_notracks}.

\acknowledgments

We thank Andy McWilliam, Chris Sneden and  Friedel Thielemann 
for helpful discussions. 
This work has been supported by  
the National Science Foundation
through grants 
AST 08-06577 to RBCH and   
AST 07-07447 to JJC.

\begin{deluxetable}{lcccccccc}
\rotate
\tablewidth{0pt}
\tabletypesize{\tiny}
\tablecaption{Observational Characteristics of the Selected Abundance Analyses\label{sobeck}}

\tablehead{
\colhead{Study}                                             &
\colhead{Telescope/Instrument\tablenotemark{a}}             &                           
\colhead{Program Stars}                                     &
\colhead{Wavelength Coverage Region [\AA]\tablenotemark{b}} &
\colhead{S/N Range}                              &
\colhead{Resolution}                             &
\colhead{Metallicity Range}                      &
\colhead{Species Detected}                       \\

}
\startdata
Gratton $\&$ Sneden 1991       & CAT/CES               & 19    & 4122-6434   & 150    & 50000  & -2.7$<$[Fe/H]$<$-0.2 & Ti I/II;V I/II;Cr I/II;Mn I;Fe I/II;Co I;Ni I\\
McWilliam et al. 1995          & LCO/echelle           & 33    & 3600-7600   & 12-47  & 22000  & -4.1$<$[Fe/H]$<$-1.9 & Ti I/II;V I/II;Cr I;Mn I;Fe I/II;Co I;Ni I\\
Feltzing $\&$ Gustafsson 1998  & McD/2-d coud\'{e}     & 47    & 3900-7000   & 200    & 100000 & -0.1$<$[Fe/H]$<$+0.5 & Ti I;V I/II;Cr I/II;Mn I;Fe I/II;Co I;Ni I\\
Cayrel et al. 2004             & VLT/UVES              & 70    & 3300-10000  & 95-430 & 45000  & -4.1$<$[Fe/H]$<$-2.7 & Ti I/II;Cr I;Mn I;Fe I/II;Co I;Ni I\\
Sobeck et al. 2006             & McD/2-d coud\'{e}; Keck/HIRES & 216  & 6010-6025 & 40-240 & 40000-60000 &-2.7$<$[Fe/H]$<$+0.1 & Mn I\\
Lai et al. 2008                & Keck/HIRES            & 28    & 3020-7665   & 8-425  & 40000  & -4.2$<$[Fe/H]$<$-2.6 & Ti I/II;V I/II;Cr I/II;Mn I/II;Fe I/II;Co I;Ni I \\

\enddata

\tablenotetext{a}{These are the primary telescope/instrument combinations employed.}
\tablenotetext{b}{For Gratton \& Sneden 1991 and Feltzing \& Gustafsson, there is only intermittent coverage in the specified wavelength range.}

\end{deluxetable}

\begin{deluxetable}{lr}
\tablewidth{0pc}
\tablecaption{Basic Input Parameters for Chemical Evolution Models\tablenotemark{1} \label{parameters}}
\tablehead{\colhead{Parameter} &\colhead{Value}}
\startdata
Star Formation Efficiency & 2.8 Gyr$^{-1}$ \\
Star Formation Power Law Exponent & 1.5 \\
Initial Mass Function Exponent, $\alpha$ & -1.20 \\
Stellar Mass Range Lower Limit & 0.08~M$_{\odot}$ \\
Stellar Mass Range Upper Limit & 40~M$_{\odot}$ \\
Infall Time Scale & 4 Gyr \\
Current Total Mass Density & 50 M$_{\odot}$ pc$^{-2}$ \\
Current Age & 13 Gyr \\
Type Ia Factor, c & 0.007 \\
Hypernova Fraction, $\epsilon$ & 0.5
\enddata
\tablenotetext{1}{Readers are referred to the appendix of \citet{HP07} for a detailed discussion of the chemical evolution code to which these parameters apply.}
\end{deluxetable}

\begin{deluxetable}{lr}
\tablewidth{0pc}
\tablecaption{WW95 Yields: Isotopes Contributing to Stable
Element\tablenotemark{1} \label{isotopes}}
\tablehead{\colhead{Element} &\colhead{Isotopes}}
\startdata
Ti & $^{46}$Ti, $^{47}$Ti, {\bf $^{48}$Ti}, $^{49}$Ti, $^{50}$Ti, $^{47}$V,
$^{48}$V, $^{49}$V, $^{48}$Cr, $^{49}$Cr  \\
V & $^{50}$V, {\bf $^{51}$V}, $^{51}$Cr, $^{51}$Mn  \\
Cr & $^{50}$Cr, {\bf $^{52}$Cr}, $^{53}$Cr, $^{54}$Cr, $^{52}$Mn, $^{53}$Mn,
$^{54}$Mn, $^{52}$Fe\\
Mn & {\bf $^{55}$Mn}, $^{55}$Fe, $^{55}$Co\\
Fe & $^{54}$Fe, {\bf $^{56}$Fe}, $^{57}$Fe, $^{58}$Fe, $^{56}$Co, $^{57}$Co,
$^{58}$Co, $^{56}$Ni\\
Co &$^{59}$Fe, {\bf $^{59}$Co}, $^{59}$Ni, $^{59}$Cu \\
Ni & $^{60}$Co, $^{61}$Co, {\bf $^{58}$Ni}, $^{60}$Ni, $^{61}$Ni, $^{62}$Ni,
$^{64}$Ni, $^{60}$Cu
\enddata
\tablenotetext{1}{The dominant stable isotope for each element in solar
system material is indicated with bold type.}
\end{deluxetable}

\begin{deluxetable}{lrr}
\tablewidth{0pc}
\tablecaption{Scaling Factors for the Kobyashi et al. Yields\label{scalingfactors}}
\tablehead{\colhead{Element} &\colhead{[Fe/H] Range}&\colhead{Scaling
Factor\tablenotemark{1}}}
\startdata
Ti  & full range & 0.15-0.063x\\
V  & x$\le$-2 & -0.34-0.32x\\
V & x$>$-2 & +0.30 \\
Cr & x$\le$-2 & +0.50+0.30x\\
Cr & x$>$-2 & -0.10 \\
Mn & x$\le$-1.5 & +0.36+0.18x \\
Mn & x$>$-1.5 & +0.10 \\
Co  & x$\le$-1.75 & -0.39-0.89x\\
Co & x$>$-1.75 & 0.0 \\
Ni  & x$\le$-1.0 & -0.75-x\\
Ni & x$>$-1.0 & +0.25
\enddata
\tablenotetext{1}{$\log_{10}$ of the scaling factor, where x=[Fe/H]}
\end{deluxetable}

\begin{deluxetable}{lcccccccc}
\tablewidth{0pc}
\tablecaption{Integrated Yields \label{integratedyields}}
\tablehead{\colhead{$\log Z$} &\colhead{[Z]}&\colhead{$\log P_{Ti}$} & \colhead{$\log P_V$} & 
\colhead{$\log P_{Cr}$} & \colhead{$\log P_{Mn}$} &  \colhead{$\log P_{Fe}$} & \colhead{$\log P_{Co}$} & \colhead{$\log P_{Ni}$}}
\startdata
-$\infty$&-$\infty$&-5.70&-5.83&-6.32&-6.52&-3.32&-5.34&-4.69 \\
-6.0&-4.3&-5.76&-6.14&-6.03&-6.34&-3.30&-5.58&-4.71 \\
-4.0&-2.3&-5.86&-6.67&-5.51&-6.04&-3.23&-5.94&-4.69 \\
-3.0&-1.3&-5.89&-6.83&-5.32&-5.87&-3.15&-6.05&-4.61 \\
-2.0&-0.3&-5.91&-5.83&-5.33&-5.70&-3.12&-5.95&-4.76 \\
-1.7&0.0&-5.88&-6.83&-5.32&-5.62&-3.13&-5.78&-4.96 \\
-1.3&+0.4&-5.88&-6.82&-5.31&-5.61&-3.11&-5.76&-5.44 
\enddata
\end{deluxetable}

\end{document}